\documentstyle[12pt]{article}    
\begin{document}           
\baselineskip=0.166667in
\begin{quote} \raggedleft TAUP 2686-2001
\end{quote}
\vglue 0.5in
\begin{center}{\bf Remarks on Superluminal Propagation \\
of an Electromagnetic Wave in Air}
\end{center} 
\begin{center}E. Comay 
\end{center}
 
\begin{center}
School of Physics and Astronomy \\
Raymond and Beverly Sackler Faculty \\
of Exact Sciences \\
Tel Aviv University \\
Tel Aviv 69978 \\
Israel
\end{center}
\vglue 0.5in
\vglue 0.5in
\noindent
PACS No: 03.30.+p, 03.50.De
\vglue 0.7in
\noindent
Abstract:

   The significance of the Lorentz gauge as a constraint on solutions
of Maxwellian waves is discussed. It is proved that recent claims 
of superluminal propagation of electromagnetic waves rely on an
erroneous basis.

\newpage
   A report on results of an electromagnetic experiment claims
that it demonstrates a superluminal speed of 
an electromagnetic signal propagating in air[1].
The experiment relies on the idea that a 
localized diffraction-free
electromagnetic wave can be created[2]. The radiation
takes the form of a cylindrically symmetric
($\varphi $-invariant) wave where the $r$-dependence is assumed
to be proportional to the Bessel function of the first kind $J_0$.
Reference [2] has arisen a great interest in utilization of $J_0$
waves and has been cited more than 340 times.
Related publication supporting the claims 
of [1] have been published[3-7]. 
Discussions of these papers and objections to the
interpretations of the experimental
results have already been published too[8-10].
The main purpose of this work is to show that one {\em cannot}
construct a diffraction-free localized electromagnetic beam whose
fields do not depend on the cylindrical coordinate $\varphi $. This
outcome proves that results of papers discussing this topic, in
general, and those ascribing superluminal velocity to beams that
take the form of Bessel function $J_0$, in particular, should
be reevaluated. Specific properties of the experiment reported
in [1] are discussed first. Later, some more general properties
of electromagnetic radiation are examined.
In this work, units where the speed of light 
$c=1$ are used.

   The experiment reported in [1] uses a cylindrically symmetric
device where electromagnetic radiation of $\lambda = 3.5\;\, cm$ is
investigated. The radiation is emitted from a horn antenna,
passes through a circular slit which is placed in the $(x,y)$ plane
and its center is on the $z$-axis. After passing the slit, 
the wave is reflected from a spherical mirror whose axis coincides with
the $z$-axis and the circular slit is placed at its 
focal plane. The electromagnetic wave 
is detected by a device lying on the
$z$-axis. The cornerstone of [1] is the following relation representing
the electromagnetic wave. It is written in cylindrical coordinates
$(\rho,\varphi,z)$
\begin{equation}
J_0(\rho k\sin \theta)exp[i(kz\cos \theta - \omega t)],
\label{eq:J0}
\end{equation}
where $J_0$ is the zeroth-order Bessel function of the first
kind, $k=\omega /c$ is the wave number and $\theta$ is a parameter.
(In what follows, Cartesian or cylindrical coordinates are
used, as required.)
As stated in [1], expression $(\!\!~\ref{eq:J0})$ is $\varphi $-invariant.

   The following argument proves that $(\!\!~\ref{eq:J0})$ cannot
describe cylindrically symmetric electromagnetic radiation
as created in the experiment described in [1]. 
Indeed, in order not to destroy
the $\varphi $-invariance, the antenna must be placed on the
$z$-axis and {\em parallel} to it. Thus, the entire device
used in [1] is invariant under a rotation around the $z$-axis. Hence,
all vectors perpendicular to this axis which are associated with
points lying on it, {\em must vanish}. Now, in electromagnetic
radiation, electric and magnetic fields are perpendicular to the
direction of propagation[11,12]. Thus, the
electric field $\bf E$ and the magnetic field $\bf B$ vanish
on the $z$-axis 
\begin{equation}
{\bf E}(\rho = 0)=0;\;\;{\bf B}(\rho = 0)=0. 
\label{eq:E0B0}
\end{equation}
Therefore, the Poynting vector, representing
the electromagnetic energy current ([11],[12], p. 237) vanishes too
\begin{equation}
{\bf S}(\rho = 0) = \frac {1}{4\pi}{\bf {E\times B}} = 0.
\label{eq:POYNTING}
\end{equation}

This argument proves that on the $z$-axis all fields interfere 
destructively. Hence, electromagnetic properties of the experiment
cannot be represented by $(\!\!~\ref{eq:J0})$, because 
on the $z$-axis $\rho = 0$ but $J_0 (0) = 1$[13]. Using continuity
properties of solutions of Maxwell equations in 
the vacuum, one infers that
electromagnetic fields near the $z$-axis are small and that the
variation from a destructive interference to a constructive
one depends on the wave length $\lambda $ and on geometrical
properties of the experiment.

   On the other hand, if, in the experiment reported in [1], 
the horn antenna emitting the radiation
is perpendicular to the $z$-axis then the radiation is not
$\varphi$-invariant(see [12], pp. 401-403, 763-767), 
contrary to what is reported in [1].

   Let us turn to the general case and prove the following theorem.
\newline
\noindent
{\em Theorem:} One cannot create a diffraction-free localized
electromagnetic beam where the intensity of the fields does not
depend on the cylindrical coordinate $\varphi$.

\noindent
{\em Proof:} Let us examine a diffraction-free
 electromagnetic radiation propagating parallel
to the $z$-axis and assume that the fields do not depend on the
cylindrical coordinate $\varphi $. As explained above, the 
$z$-components of the electric and the magnetic field vanish. Hence,
the following Maxwell equation in the vacuum takes the form
\begin{equation}
\nabla \cdot {\bf E} = \frac {\partial E_x}{\partial x} +
\frac {\partial E_y}{\partial y} = 0.
\label{eq:MAXE1}
\end{equation}
Let us examine this equation at a point $P_1=(x\neq 0,0,0)$. Using
the assumed $\varphi $-invariance of the 
field components and the following
expression for the cylindrical coordinate $\rho = (x^2 + y^2)^{1/2}$,
one obtains from $(\!\!~\ref{eq:MAXE1})$
\begin{equation}
\frac {\partial E_x}{\partial \rho} \frac {x}{(x^2 + y^2)^{1/2}} +
\frac {\partial E_y}{\partial \rho} \frac {y}{(x^2 + y^2)^{1/2}} = 0.
\label{eq:MAXE2}
\end{equation}
Now, at point $P_1$, $y=0$ and $x\neq 0$. Hence 
\begin{equation}
\frac {\partial E_x}{\partial \rho} = 0.
\label{eq:DEX}
\end{equation}

   Examining the system at
another point $P_2 = (0,y\neq 0,0)$, which lies on the
$y$-axis, one obtains from a similar procedure
\begin{equation}
\frac {\partial E_y}{\partial \rho} = 0.
\label{eq:DEY}
\end{equation}

   Results $(\!\!~\ref{eq:DEX})$ and $(\!\!~\ref{eq:DEY})$ hold
for all values of $\rho$, except $\rho = 0$. Continuity properties 
of the derivative of the field components in vacuum prove that 
$(\!\!~\ref{eq:DEX})$ and $(\!\!~\ref{eq:DEY})$ hold for $\rho = 0$
too. The null derivative of the electric field components means
that they equal a constant in the entire $(x,y)$ plane.
It can be shown analogously that the magnetic field $\bf B$
shares the same properties.

   This analysis proves that if the radiation is diffraction-free and
$\varphi $-invariant then the electric and the magnetic fields
are uniform
on the entire $(x,y)$ plane. Hence, the radiation is the well
known plane wave (see [11], pp. 110-112, [12], pp. 269-273), 
which is {\em not} localized (because it does not
vanish for $x=\infty,\;\; y=\infty$). This 
conclusion completes the proof.

   It follows from this theorem that electromagnetic fields cannot
take the diffraction-free form of the Bessel function
$(\!\!~\ref{eq:J0})$.

The results obtained above deserve a further discussion.
It is well known that, in the vacuum, Maxwell equations yield 
the homogeneous wave equation
for components of the electromagnetic field.
Some authors obtain the homogeneous wave equation for the field
components directly from Maxwell equations in the vacuum[14,15].
Hence, satisfying the homogeneous
wave equation is a {\em necessary} condition for being consistent with
Maxwell equation in the vacuum. 
At this point, the problem whether the condition
is {\em sufficient} is not yet decided.

   This issue can be treated alternatively (see [11], pp. 108-109, 
[12], pp. 219-220, [14], p. 67, [15] pp. 313-314 and [16,17]). 
Indeed, an analysis of the
4-potentials $ A^{\mu } = (\Phi,{\bf A})$ proves that Maxwell 
equations are equivalent to the inhomogeneous wave equations
\begin{equation}
\Box A^{\mu } = 4\pi j^{\mu },
\label{eq:INHOMOGENEOUS}
\end{equation}
provided the Lorentz gauge 
\begin{equation}
\frac {\partial \Phi}{\partial t} + \nabla \cdot {\bf A} = 0
\label{eq:LORENTZ}
\end{equation}
holds. This outcome indicates that satisfying the wave equation alone
is not sufficient for being consistent with Maxwell equations.

   As a matter of fact, it can be proved in a very simple way that 
vector fields whose components
satisfy the homogeneous wave equation in the vacuum
do not necessarily satisfy Maxwell equation.
Indeed, assume that each of the vector field components
$(E_x,E_y,E_z)$ satisfies in the vacuum
the homogeneous wave equation
\begin{equation}
\Box E_x = 0;\;\;\;
\Box E_y = 0;\;\;\;
\Box E_z = 0.
\label{eq:WAVEE}
\end{equation}
Now, not all field components vanish throughout the entire space.
Let $E_x$ be nonzero in a certain region. As a physical 
field, it vanishes at infinity. Therefore, $E_x$
is not a uniform function of $x$ and its derivative is
nonzero somewhere 
\begin{equation}
\frac {\partial E_x}{\partial x} = f(x,y,z) \neq 0.
\label{eq:DEXDX}
\end{equation}
Due to Maxwell equation in the vacuum
\begin{equation}
\nabla \cdot {\bf E} = 0,
\label{eq:DIVE}
\end{equation}
one must have at this region  
\begin{equation}
\frac {\partial E_y}{\partial y} + \frac {\partial E_z}{\partial z} = 
-f(x,y,z).
\label{eq:DEYZDYZ}
\end{equation}
This result provides an example of field components $(E_x,E_y,E_z)$
that satisfy the homogeneous wave equation 
$(\!\!~\ref{eq:WAVEE})$ {\em and} Maxwell equation 
$(\!\!~\ref{eq:DIVE})$ as well.
However, one may use another system of fields whose components are
$(2E_x,E_y,E_z)$. Here, $E_x$ of the previous system is multiplied
by 2. Obviously, the new field components do satisfy 
the homogeneous wave equation $(\!\!~\ref{eq:WAVEE})$
but violate Maxwell equation $(\!\!~\ref{eq:DIVE})$. 
This conclusion completes the proof.

The quantities $(E_x,E_y,E_z)$ used above may represent the
electric field as well as the magnetic one.

   This work shows the significance of the Lorentz gauge
$(\!\!~\ref{eq:LORENTZ})$ constraint which is imposed on
the 4-potentials $(\Phi,{\bf A})$ 
which are solutions of the wave equation.
In particular, one should not accept automatically field components
that solve the homogeneous wave equation in the vacuum as 
solutions of Maxwell equations.
An outcome of the discussion carried out here is that the Bessel
function $(\!\!~\ref{eq:J0})$ suggested in [2]
and used in [1] for an apparent
demonstration of superluminal properties of electromagnetic radiation
contains a fundamental error because it is just a special case
of a localized diffraction-free $\varphi $-invariant beam. Therefore,
results of theoretical and experimental works[1-9] relying on 
$(\!\!~\ref{eq:J0})$ should be reevaluated. Obviously, special care
must be taken in an evaluation of claims concerning 
the existence of superluminal properties of electromagnetic
radiation, a
claim which is inconsistent with special relativity and
Maxwellian electrodynamics in the vacuum.

\newpage
References:
\begin{itemize}
\item[{*}] Email: eli@tauphy.tau.ac.il
\item[{[1]}] D. Mugnai, A. Ranfagni and R. Ruggeri, Phys. Rev.
Lett. {\bf 84}, 4830 (2000).
\item[{[2]}] J. Durnin, J. J. Miceli, Jr. and J. H. Eberly,
Phys. Rev. Lett. {\bf 58}, 1499 (1987).
\item[{[3]}] D. Mugnai, Phys. Lett. {\bf 278A}, 6 (2000).
\item[{[4]}] M. Zamboni-Rached and D. Mugnai, Phys. Lett. 
{\bf 284A}, 294 (2001).
\item[{[5]}] D. Mugnai, Phys. Lett. {\bf 284A}, 304 (2001).
\item[{[6]}] A.M. Shaarawi and I. M. Besieris, J. of Phys. 
{\bf A33}, 7227 (2000).
\item[{[7]}] A.M. Shaarawi and I. M. Besieris, J. of Phys. 
{\bf A33}, 7255 (2000).
\item[{[8]}] W. A. Rodrigues Jr, D. S. Thober and A. L. Xavier Jr,
Phys. Lett. {\bf 284}, 217 (2001).
\item[{[9]}] E. Capelas de Oliveira, W. A. Rodrigues Jr, D. S. Thober
and A. L. Xavier Jr, Phys. Lett. {\bf 284}, 296 (2001).
\item[{[10]}] Thilo Sauter and Fritz Paschke, Phys. Lett. {\bf 285}, 1
(2001).
\item[{[11]}] L. D. Landau and E. M. Lifshitz, {\em The Classical
Theory of Fields} (Pergamon, Oxford, 1975). P. 111.
\item[{[12]}] J. D. Jackson, {\em Classical Electrodynamics} (John Wiley, 
New York,1975). p. 395. 
\item[{[13]}] M. Abramowitz and I. Stegun, {\em Handbook of
Mathematical Functions}, (Dover, New York, 1965). p. 360.
\item[{[14]}] F. Rohrlich, {\em Classical Charged Particles},
(Addison-wesley, Reading mass, 1965). p. 63.
\item[{[15]}] J. R. Reitz and F. J. Milford, {\em Foundations of
Electromagnetic Theory} (Addison-Wesley, Reading, 1967) p. 300.
\item[{[16]}] R. P. Feynman, R. B. Leighton and M. Sands, {\em The Feynman
Lectures on Physics} (Addison-Wesley, Reading, 1964). Vol. 2. pp.
18-11, 21-2.
\item[{[17]}] H. A. Haus and J. R. Melcher, {\em Electromagnetic Fields
and Energy} (Prentice Hall, Englewood Cliffs, 1989). pp. 538-539.
\end{itemize}

\end{document}